\documentclass{PoS}

\title{Accretion in Active Galactic Nuclei}

\ShortTitle{Accretion in Active Galactic Nuclei}

\author{\speaker{Bozena Czerny}\\
        Center for Theoretical Physics, Polish Academy of Sciences, Al. Lotnikow 32/46, Warsaw, Poland, P.O.Box: 02-668\\
        E-mail: \email{bcz@cft.edu.pl}}

\author{Mohammad-Hassan Naddaf\\
        Center for Theoretical Physics, Polish Academy of Sciences, Al. Lotnikow 32/46, Warsaw, Poland, P.O.Box: 02-668\\
        Nicolaus Copernicus Astronomical Center, Polish Academy of Sciences, Ul. Bartycka 18, Warsaw, Poland, P.O.Box: 00-716\\
        E-mail: \email{naddaf@cft.edu.pl}}

\abstract{We review the current status of the understanding how the accretion onto the central black hole proceeds in Active Galaxies. Standard accretion disk is a key element in all relatively bright active galaxies like Seyferts and quasars, although it is not present in very low luminosity sources, like Sgr A*. However, the standard disk does not explain the broad band spectrum, so the disk has to be supplemented with a number of additional components, and our deeper understanding of these components is still far from being complete. These additional elements are: compact hard X-ray corona, inner hot flow, warm corona, disk wind and the Broad Line Region, and finally dusty/molecular torus. All these elements seem to be needed in various proportions, depending predominantly on the Eddington ratio of a given source. These elements also interact with each other which is not yet fully taken into account.}

\FullConference{Accretion Processes in Cosmic Sources - II - APCS2018\\
		3-8 September 2018\\
		Saint Petersburg, Russian Federation}

\begin{document}

\section{Introduction}

Active Galactic Nuclei are now basically understood, and we accept a universal picture of them, whether the nucleus is only weakly active, like Sgr A* at the center of the Milky Way, or strongly active as distant quasars. The nucleus of every galaxy contains a supermassive black hole. The level of nucleus activity depends on the amount of matter inflowing onto the black hole which varies over time. As the matter inflows, it loses angular momentum and energy, and the emission of matter from close vicinity of black hole provides us with information about this spatially unresolved region.

Despite the general agreement about the underlying mechanism of the emission, the progress in understanding the details of the process is rather slow. This is because the nucleus region is actually quite messy, with outflow accompanying inflow, and the atmosphere of active nucleus is partially surrounded by dynamically changing clouds.

\begin{figure}[t]
\centering
\includegraphics[scale=0.7]{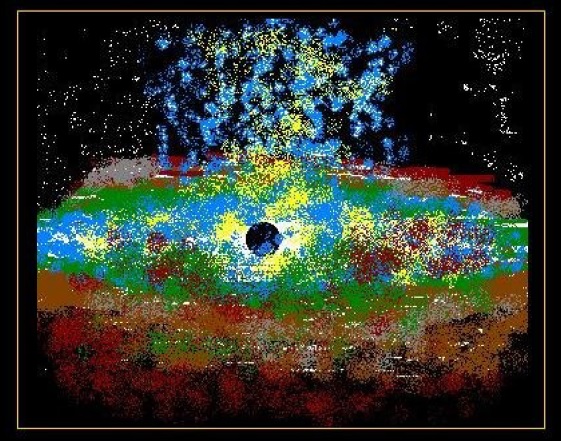}
\caption{A schematic view of the messy central region by authors; for more elegant drawings see \cite{UrryPadovani}.}
\label{fig:MessyCentralRegion}
\end{figure}

\section{Standard Accretion Disk}

A firm element of theoretical explanation of the AGN activity is the standard model of an accretion disk, firstly introduced by \cite{Shakura}, and then fully developed by \cite{ShakuraSunyaev}.
The Keplerian disk accretion has been earlier introduced by \cite{Lynden} in the context of AGN but it lacked the power of parameterization by viscosity parameter $\alpha$.
Relativistic version of the standard model was later presented by \cite{NovikovThorne}.

\begin{figure}[t]
\centering
\includegraphics[scale=0.5]{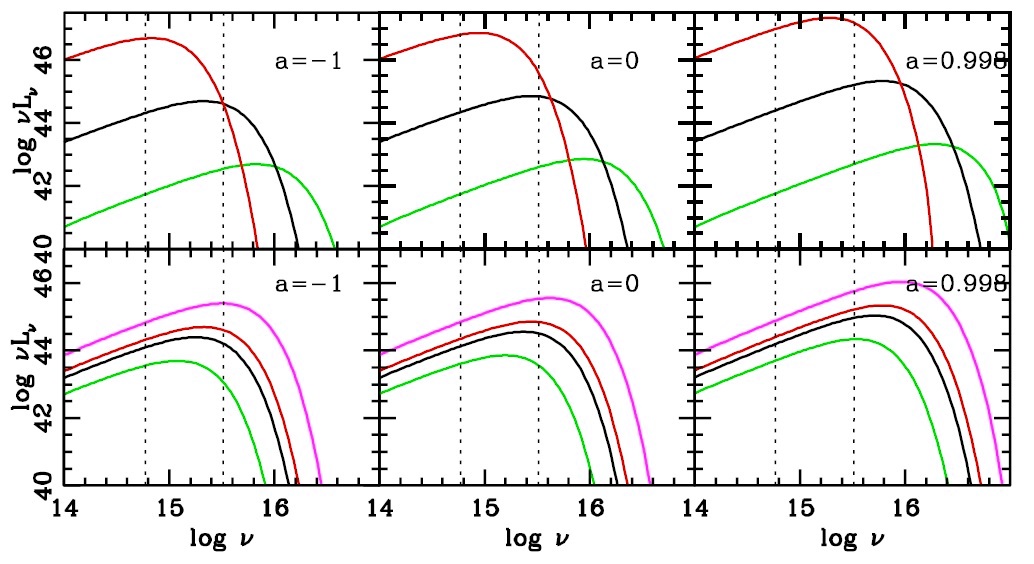}
\caption{The exemplary set of spectra from standard accretion disks for a range of spins. Upper panels show the dependence on black hole mass ($10^6, 10^8$ and $10^{10} M_{\odot}$, fixed $L/L_{Edd}=0.1$). Lower panels show the dependence on Eddington ratio (0.01, 0.05, 0.1 and 0.5, for a fixed mass of $10^8 M_{\odot}$). The figure is adopted from \cite{Czerny2019}.}
\label{fig:Spectra}
\end{figure}

The model in its most basic form is very simple. It is assumed that the accreting matter has a very large angular momentum, viscous forces are responsible for the transfer of angular momentum outwards, and a stationary inflow with the accretion rate $\dot M$ slowly proceeds onto the black hole through a set of Keplerian circular orbits, with zero-torque boundary condition at the disk inner radius, usually located at Innermost Stable Circular Orbit (ISCO). The disk radiates locally in the form of a black body. Knowing the black hole mass, $M_{BH}$, its spin, $a$ (in full GR case), and  $\dot M$, one can determine the flux as a function of radius and also the temperature profile of the disk, uniquely and independently of the viscosity mechanism. Inner boundary condition sets the maximum value of the accretion disk temperature. Assuming that disk is locally emitting as a black body, we can calculate the accretion disk spectra for a range of black hole masses, Eddington rations and spins. Examples of such theoretical spectra are shown in Fig. \ref{fig:Spectra}.

Such theoretical spectrum resulted from the standard model of accretion disk is frequently seen in the spectra of quasars. For example, authors in \cite{Capellupoetal} showed several quasar spectra all nicely fitted by this simple model, and only some sources required correction for the reddening in the host galaxy. However, the presence of this element is a characteristic feature only for those sources efficiently accreting with Eddington luminosity ranging between a few percent and below 0.3.

\section{Elements of the puzzle}

For quasars, this is not yet the end of story, but the beginning. The broad band spectra of quasars are far more complex (see Fig. 11 in \cite{Richardsetal}). Optical-UV band is indeed dominated by the Big Blue Bump which is well described by the classical accretion disk model, but additionally we see considerable amount of emission in IR, which comes from the hot dust around the nucleus (dusty/molecular torus) and the host galaxy. In the soft X-rays there is a relatively steep power law, and in hard X-rays another power law emission is usually seen, extending up to $\sim 100$ keV. Blazars (radio-loud objects) have an additional strong jet which is oriented toward an observer, and can dominate the entire broad band spectrum due to the Doppler boosting. In my talk I will concentrate on the radio-quiet sources, leaving the jetted sources aside.

In order to model the entire broad band spectrum, including the variability, and to well represent all types of active nuclei, we need to have additional elements of the puzzle which do not come from the first principles. Therefore, these elements are considered to be there, and are then tailored with respect to some basic physical properties. We are not yet at a position to say that assuming some inflow of matter at the outer radius would reproduce all these elements. Such computations would need proper treatment of the magnetic field, radiative transfer (including ionization state), dynamics - all in a very broad range of timescales. Thus what is usually done is to invent these additional elements and look for observational constraints on their properties.

Thus, apart from the disk, we usually consider the following elements which I will discuss them below:

\begin{itemize}
\item compact corona/jet base
\item inner hot flow
\item warm corona
\item disk wind/Broad Line Region
\item dusty/molecular torus
\end{itemize}

\subsection{compact corona/jet base/lamp-post}\label{sec:compactcorona}

This element is most often known as lamp-post model since it is sometimes technically assumed to be a point-like sources of hard X-rays. Historically, it can be connected to the jet-base model already proposed by \cite{HenriPelletier}. This element is supposed to be responsible for the hard X-ray emission.

Observationally, such element is present both in high and low Eddington ratio sources, although its contribution to the total energy budget in high Eddington sources is small. Kubota \& Done proposed that the total luminosity of the component saturates at 2\% of the Eddington flux \cite{KubotaDone}, which might be related to the raising efficiency of the pair creation \cite{BisnovatyiKogan, Svenssonetal82, Svenssonetal83, Svenssonetal84, Zdziarski, Fabianetal1986}. Also magnetic field reconnection is likely to play the important role there, e.g. \cite{Beloborodov2017}.

There are observational constraints on the size of this region, $1 - 10 R_{Schw}$, which come from variability \cite{Fabianetal}, X-ray reflection \cite{DeMarcoetal, Karaetal}, and quasar microlensing in X-ray band \cite{Mosqueraetal}. The jet formation, and perhaps the compact corona under discussion, requires the presence of the large scale magnetic field, which in turn requires geometrically thick inner disk. The important process in this region is also magnetic field reconnection. It has probably a close connection to the concept of blobby ejection which is a characteristic feature of high Eddington sources (e.g. 3C 273, \cite{Courvoisier}). It is interesting to note that rapid flares in the most inactive AGN, Sgr A*, are also well-described by a ball of order of $R_{Schw}$ in size \cite{Markoffetal}. Another physical process which might be closely relevant is the idea of Magnetically Arrested Disks (MAD) firstly constructed by \cite{Ruzmaikin1976}, then rediscovered after 27 years in
\cite{Narayanetal}. Full understanding of this region is still far. However, the issue takes us to the problem of the inner disk. 

\subsection{Inner Hot Flow}

Inner Hot Flow is still the matter of the debate but there are two important reasons why it has to develop in the innermost part of the disk. The first reason is that jets need a large scale magnetic field to be formed, and such a large scale magnetic field can be supported by a geometrically thick accretion flow. The second reason is that the existence of the cold Keplerian Shakura-Sunyaev accretion disk in very low Eddington ratio flows, e.g. Sgr A*, is firmly excluded so the accretion must proceed in the form of geometrically thick hot flow. This suggests that the extension of the hot flow depends strongly on the source Eddington ratio. However, it does not set the flow nature.

There are two basic geometrical setups for such a flow: radial and vertical. The first one is well known since many years as ADAF (Advection-Dominated Accretion Flow), firstly proposed by \cite{Ichimaru}, which became popular later on, starting with a paper by \cite{NarayanYi}. ADAF flow is widely different from the Shakura-Sunyaev solution: the inflow is fast; the angular momentum of the plasma is only a fraction of the local Keplerian angular momentum; the plasma is two-temperature, with ion temperature close to the local virial temperature; the dissipated energy is transferred from ions to electrons which are cooled down efficiently by Comptonization of the soft photons, which either come from the outer cold disk or are locally generated by the synchrotron mechanism. In this scenario, we have in general an outer cold disk and an inner hot flow. In the second scenario, we have a hot flow above the cold disk, and such solutions were proposed by \cite{ChakrabartiTitarchuk, Zyckietal}. It can be named as coronal flow. The physics of the hot phase is similar to the previous case, but the overlap with the cold disk makes the presence of the soft photons more important. In reality, we can likely have a combination of the two, sandwich-type coronal flow in the outer part and a purely hot flow in the innermost part. Coronal flow has been studied in a number of papers. Some of them discussed the physics of the mass exchange between the disk and the corona, which results from the heating/cooling balance in the two media plus electron conduction between the disk and corona \cite{RozanskaCzerny, LiuetalMeyer, Liuetal2007, Meyeretal, Liuetal2002}. Time-dependent two-phase flow has been modeled by \cite{Mayer}, showing that at high accretion rates the corona flow is weak but at low accretion rates the disk is evaporated and a pure hot flow forms. The strong dependence of the cooling efficiency on density leads us to a prediction that actually, at medium accretion rates, an inner cold disk can form, separated from the outer cold disk by a gap made of hot flow only. 
The full physics of the hot flow is still not fully understood. ADAF model cannot be formed in presence of equipartition between magnetic and kinetic energy, what was suggested in the model, because of heating due to magnetic field reconnection \cite{Lovelace1997}, and this reconnection process studies accelerated only in the recent years.

Observationally, the issue is not so easy to set. Determination of the transition radius between the outer cold disk and the inner hot flow is not unique even for the same data sets \cite{Garciaetal2015}.

\begin{figure}
\centering
\includegraphics[scale=0.5]{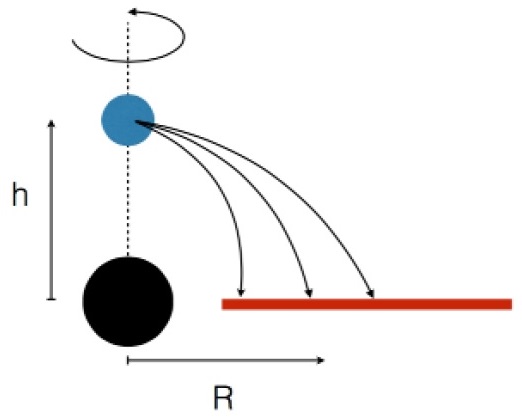}
\caption{The schematic view of the inner cold disk and a compact corona. Hard X-ray emission from the corona irradiates the disk. Mass supply to the hot corona is not included in this picture.}
\label{fig:corona}
\end{figure}

The presence of the hot flow should have some observational consequences. The hot flow should provide the mass to the hot compact corona, so this medium should be between the inner cold disk, if it is there, and an observer. This is never discussed in the simple parametric models (see Fig. \ref{fig:corona}). 

Hot flows easily lead to some outflow as already discussed by \cite{Blandford} who introduced a concept of Advection Dominated Inflow-Outflow Solutions (ADIOS). This outflow takes mass, angular momentum and energy away from the main flow which then cannot be treated as conservative. We cannot predict the efficiency of such an outflow, but in some cases it can be observationally tested. For example, such an outflow can provide an explanation why in galactic black holes the value of spin, determined by the continuum fitting method apparently, depends on the luminosity state of the source \cite{Youetal}.

\subsection{Warm corona}

Hot corona around accretion disk, together with a convective
instability, was first considered in \cite{Lynden1976, Blinnikov1977}. Formation of such corona due to heating by magnetic reconnection was calculated in \cite{Galeev1979}.
After 15 years, a static corona in a sandwich geometry has been introduced by \cite{HaardtMaraschi} to explain the hard X-ray emission. However, authors in the original paper already showed that such geometrical setup does not produce hard X-ray spectra. Later patchy corona \cite{HaardtGhisellini, Sternetal1995} or outflowing corona \cite{Beloborodov1999} were introduced which aimed at obtaining harder spectra through a decrease of the soft photon flux due to geometrical or relativistic effect. Hard X-ray emission models later shifted towards the compact corona (see Sect. \ref{sec:compactcorona}). However, the need for a sandwich-type corona reappeared when more attention was payed to soft X-ray spectra and soft X-ray excess. In particular, soft X-ray excess in many AGN is well fitted by Comptonization in a medium with very moderate temperature, of order of 1 keV or less \cite{Janiuk2001, Petruccietal2018}, but large optical depth. We call this medium warm corona to differentiate from the compact hard X-ray corona discussed before. Such corona must be heated by magnetic dissipation, and cooled by Comptonization.

Such warm corona is theoretically attractive since besides reproducing soft X-ray spectra it can also stabilize the accretion disk at the same time which is otherwise subject to the radiation pressure instability, e.g. \cite{LightmanEardley, PringleRees, Czernyetal}. The existence of such temperature inversion in the optically thick medium (warm corona is hotter than the disk inside it) has been discussed by \cite{Rozanskaetal} where it was shown that the medium can have an optical depth of up to 20, being cooled by Comptonization, and still remains in hydrostatic equilibrium, if the role of the magnetic pressure is dominant. Some researchers proposed a detailed model of how the magnetic heating of such corona can work \cite{BegelmanSilk}.

Warm corona is necessary to fit the broad band spectra of AGN, and such element is for example included in the new SED fitter proposed by \cite{KubotaDone}. 

\begin{figure}
\centering
\includegraphics[scale=0.35]{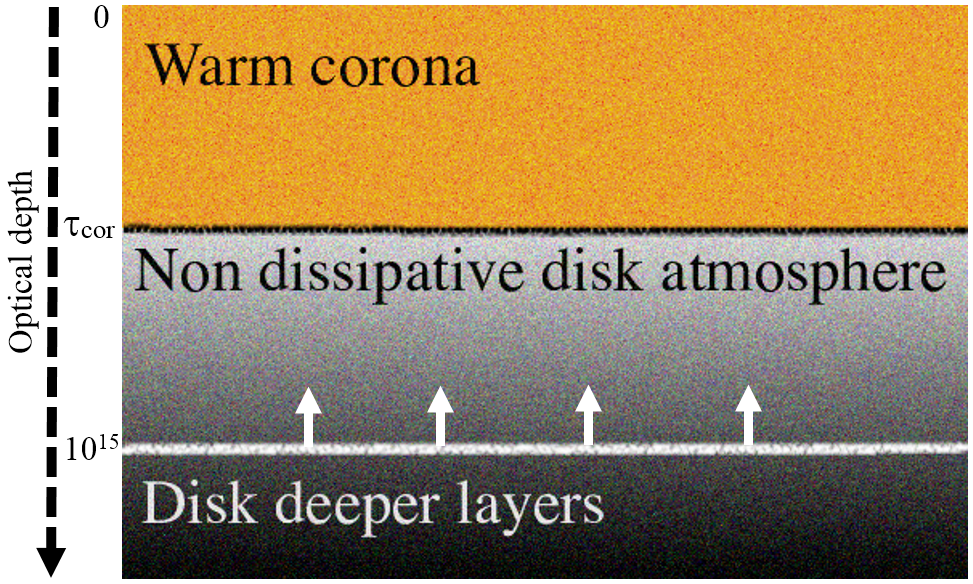}
\caption{The schematic view of the warm corona.}
\label{fig:warmcorona}
\end{figure}

\subsection{Inverse-Compton heated extended corona and wind}

Such an element has been theoretically predicted many years ago \cite{Begelmanetal}. If the radiation produced close to the black hole irradiates the outer part of a standard accretion disk, a hot layer forms above the disk, with the temperature determined by the balance between Compton heating and Compton cooling (i.e. Inverse Compton temperature). The value of the temperature depends only on the spectral shape of the incident radiation and typically is about $10^6 - 10^8$ K. Close enough such a temperature is smaller than the virial temperature, and the medium forms a static extended corona, further out the virial temperature is smaller and the heated medium inevitably forms a wind. The transition between the static corona and a wind region takes place at about $10^5 R_{Schw}$. 

Observationally, such an extended corona is observed in galactic sources, e.g. \cite{WhiteHolt, Steineretal}. In AGN the presence of this fully ionized medium is less clear. It is probably not related to Ultra-Fast Outflows (UFO) \cite{Longinotti} but perhaps this medium forms a confining medium for denser clouds of the Broad Line Region.

\subsection{Line-driven winds}

The accretion disk atmosphere in AGN disks is not as highly ionized as in galactic sources, so Compton-driven rather than line-driven wind is expected \cite{ProgaStoneKallman, Elvis2012}. Classical model of line-driven wind from stars, and other objects, was developed in \cite{Castor1975}.
Line-driven winds may well constitute the significant part of the Broad Line Region. As introduced by \cite{CollinSouffrin}, BLR basically consists of two parts: High Ionization Lines (HIL), and Low Ionization Lines (LIL). HIL is located closer in, shows lower local densities and lower column densities in comparison to LIL. HIL examples are CIV and He II lines which generally show signatures of outflow. Balmer lines as well as Fe II pseudo-continuum belong to LIL class, and the outflow signature is much fainter or absent in these lines. The connection between the line driven winds and BLR was already explored in much detail by \cite{Murrayetal, MurrayChiang}.

Observationally, there are many signatures of outflows in AGN, in the form of UFO, WA (warm absorber), Broad Absorption Line (BAL), and asymmetry in some emission lines. However, detailed undestanding of the launching process is not easy. Force multiplier due to the line absorption can be above 100, but too strong irradiation leads to over-ionization of the plasma and reduction of the radiation pressure. Thus, a need for some shielding of the launching region has been under discussion for many years, e.g. \cite{GallagherEverett}. But too much shielding would prevent launching the outflow, and the appropriate balance would also be difficult to reach in self-consistent computations transfer, e.g. \cite{Higginbottom}.

\section{Broad Line Region} 

\begin{figure}[b]
\centering
\includegraphics[scale=0.45]{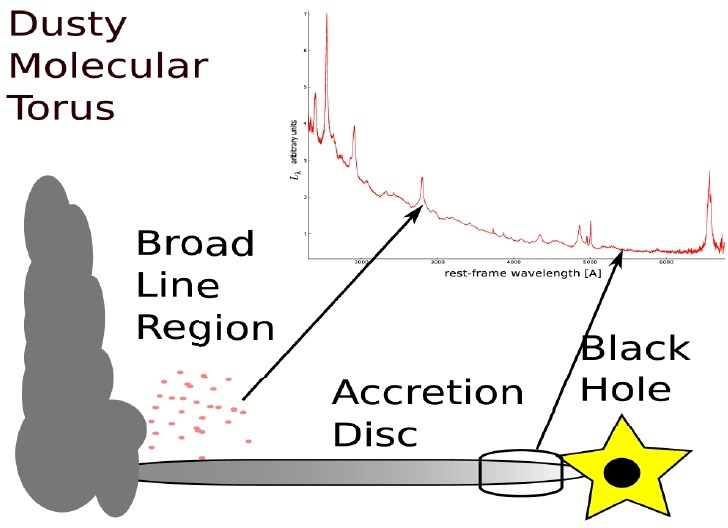}
\caption{The schematic view of disk, BLR and the dusty torus, together with the optical/UV spectrum of an AGN (with the permission from Krzysztof Hryniewicz).}
\label{fig:DustyTorus}
\end{figure}

Broad emission lines in the optical and UV band are the most characteristic features of AGN. Not all AGN have them: objects with broad lines are classified as type 1 AGN, and objects without broad lines as type 2 AGN \cite{Khachikian1971, Khachikian1974}. Type 2 AGN is the heterogeneous class: some of them have BLR hidden behind the dusty/molecular torus, and observer can see broad emission lines in polarized light scattered by extended medium \cite{AntonucciMiller}, but some of type 2 AGN are rather true type 2 AGN, devoid of the BLR which seem to belong to low luminosity AGN. Observationally, obscuration and the existence of the hidden BLR is not always easy to exclude, but sources like Sgr A* certainly do not have any BLR.

The location of the BLR can be determined by the reverberation studies. The basic measurement is the determination of time delay of the emission line with respect to the continuum \cite{Liutyi1977, Kaspietal, Bentzetal}.
The geometry of BLR, however, is not given by a single number. For example, a 2-D numerical model of BLR is calculated in \cite{Dorodnitsyn}.
There are also successful attempts to reconstruct the 3-D geometry of the BLR as an extended region, e.g. \cite{DoneKrolik, Grieretal}. 
These results show that the motion of the BLR clouds is predominantly Keplerian, but with an additional inflow or outflow component.

The simple measurement of the effective time delay is particularly important since it allows for the black hole mass determination. Time delay measurement provides the effective radius of the BLR, line width gives the velocity, and combining these two quantities we can calculate the black hole mass from the Kepler's law. There are of course some additional issues, like the use of the Full Width Half Maximum (FWHM) or $\sigma$ as the descriptor of the line shape, e.g. \cite{Collinetal}, and the issue of the proportionality coefficient which is related to the geometrical distribution of cloud orbits.

Reverberation campaigns were mostly done for relatively nearby AGN, for H$\beta$ line since this line is strong, and additionally the presence of the narrow [OIII]5007\AA~ line was convenient for calibration of results obtained from different telescopes.

Monitoring led to discovery of the tight relation between the monochromatic flux and the time delay \cite{Kaspietal, Bentzetal}. Such a relation opened a way to massive black hole mass determinations in thousands of AGN at the basis of a single spectrum measurement \cite{VestergaardPeterson, Calderoneetal}.

Theoretically, such a relation was surprising since the general expectation was that the BLR distance should scale as the square root of the ionization flux, not as the square root of the monochromatic luminosity in the optical band. However, a simple theoretical explanation of such a scaling has been formulated by \cite{CzernyHryniewicz}. In their approach the onset of the BLR is defined by the disk radius where the material is efficiently shifted above the disk by the radiation pressure acting on dust. We know well from stellar astrophysics that dusty winds are orders of magnitude more efficient than line-driven winds in hotter stars. The Failed Radiatively Accelerated Dusty Outflow (FRADO) provides a physical mechanism to rise up the material from the disk in the form of massive wind but as soon as the gas is better exposed to the irradiation by the central region the dust is evaporated, the radiation pressure support is lost and the matter falls back toward the disk. This picture is suitable for the LIL part of the BLR and explains why there are no strong outflow signatures. The model in its simple analytical form is consistent with the data and can approximately predict the line shapes \cite{Czernyetal2017}. Similar model, but static, has been recently developed by \cite{BaskinLaor}. Theoretical explanation allows us to safely extend the relation for sources with higher mass and higher redshift, and to use it both for black hole mass determination and for cosmology, as tracers of the expansion rate of the Universe \cite{Watsonetal, Haasetal, Czernyetal2013}.

\subsection{clumpiness of the BLR region}

We frequently refer to the BLR clouds, but the actual physics of the region is not yet described in time-dependent way with high precision. The disk wind starts as a roughly continuum medium, but the thermal instability operating in hard X-ray irradiated medium leads to cloud formation \cite{KrolikMcKeeTarter}. Indeed, rare cases of the X-ray eclipses with timescales of few days appropriate for BLR clouds indicate a cometary shape of a single cloud \cite{Maiolino}. The thermal instability predicts the optical depth of the cloud to be of order of $10^{23} - 10^{24}$ $cm^{-2}$.

Another way of looking at the cloud formation is in terms of the radiation pressure confinement \cite{Sternetal2014, BaskinLaor}. In this picture, a dynamical condition instead of thermal balance is superimposed. This, however, requires the pre-existence of a boundary, and does not depend on the spectral shape. 

The formation of a two-phase medium leads to a more complex dynamics, in particular a combination of inflow and outflow, depending on the medium conditions. Denser colder clouds may inflow more easily while surrounding hot plasma may have temperature higher than virial and flows out. Examples of such plasma coexistence can be found in many locations, including Sgr A* region \cite{Rozanskaetal2014, Baraietal, Elvis}. However, colder clouds may be subject of increased radiation pressure from the nucleus, so that they will rather flow out than in, like in a simulation by \cite{Baraietal2012}. Net effect will strongly depend on many details.

\section{Outer radius of the accretion disk}

Schematic pictures frequently show the accretion disk surrounding the black hole and the BLR at larger radii. However, the overlap between the BLR and the disk is not firmly settled. Observationally, in bright quasars the accretion disk extends well into the IR but this spectral component can only be seen in the polarized light \cite{Kishimotoetal}, otherwise it is buried in the starlight from the host galaxy. Its extension corresponds to disk atmosphere temperatures of about 1800 K. At longer wavelengths the hot dust is seen but this component comes from the dusty/molecular torus.

\begin{figure}[t]
\centering
\includegraphics[scale=0.39]{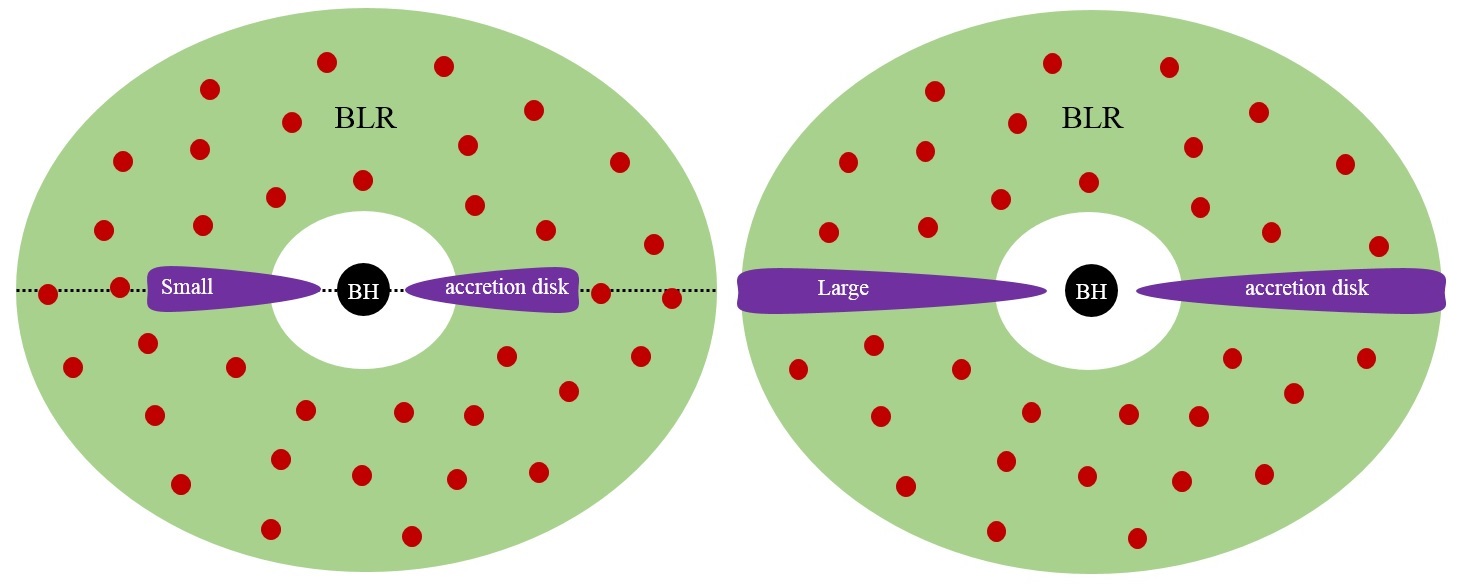}
\caption{The schematic view of accretion disk overlaping or not overlaping with BLR.}
\label{fig:DiskBLR}
\end{figure}

In FRADO or any wind-based model of the BLR the disk has to be present in the BLR region. However, BLR models based on cloud inflow from larger radii, e.g. \cite{Wangetal2017}, do not imply such a condition. Observationally, it is not easy to say whether we see clouds only above the equatorial plane (those below are shielded by the disk) or on both sides of the equatorial plane (no disk). Theoretically, there is a well known mechanism which can actually lead to efficient destruction of the outer disk: gravitation instability \cite{CollinZahn, Wangetal}. The comparison of the predicted effect of self-gravity of the disk with the location of the BLR done in \cite{Czernyetal2016} did not show any signature of the gravitational instability. Also the parameter space characteristic for true type 2 AGN (those having low Eddington ratios) is rather consistent with disappearance of the BLR due to the destruction of the cold Keplerian disk and replacement with an inner ADAF, e.g. \cite{Czernyetal2004}.

Physically, the issue of the outer disk radius is connected with the issue of the interaction of the nucleus with the host galaxy. First, we need a reservoir of material to feed the nucleus, and the angular momentum of this material along with the corresponding circularization radius gives a certain constraint. Next, since in the disk accretion flow the angular momentum is transported outside, it causes a continuous expansion of the disk outer radius unless there is an additional mechanism of angular momentum removal like magnetic wind or the action of tidal forces from the Nuclear Stellar Cluster. Observationally, we do not have the proper insight into the issue since our spatial resolution reaches an inner few parsecs only in some nearby AGN \cite{Asmusetal}.

\section{Summary}

The approach we described in this review is mostly parametric, although using physical laws for description of the specific elements of the puzzle. An alternative approach is also possible: to use just physical laws and see what we can get. The second approach is represented with 3-D MHD simulations supplied by radiative transfer, and an excellent example is a model of accretion disk in cataclysmic variables (Blaes, this volume). However, such a disk has much smaller dynamical range (basically the ratio of the outer to inner radius) than an AGN accretion disk, but it is still far from grasping the whole disk evolution and all spectral elements (X-rays, for example). Thus the progress has to be done from both sides with the hope that at some point the two approaches will meet each other and will provide us with a full picture of the accretion pattern onto black holes in general, including a case of AGN. However, we are still far from that. Thus, as a summary we want to draw your attention to two semi-global models which are worth further detailed consideration.

\begin{figure}[b]
	\centering
	\includegraphics[scale=0.3]{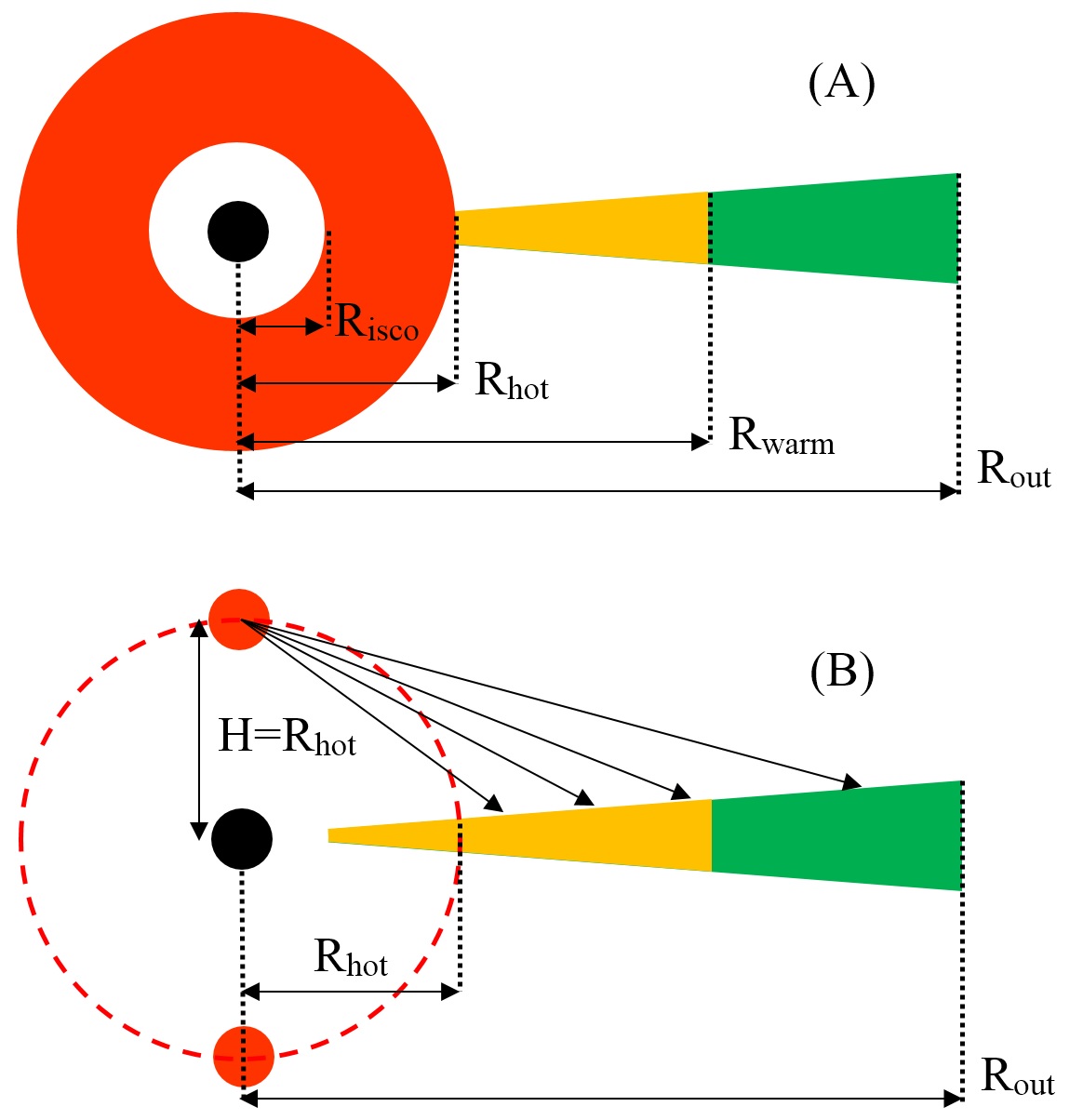}
	\caption{The schematic view of the Kubota \& Done model which provides broad band spectra \cite{KubotaDone}.}
	\label{fig:KubotaDone}
\end{figure}

\subsection{global SED model}  

New global SED model has been recently proposed by \cite{KubotaDone}. The model is parametric but it conveniently combines the three elements discussed above: hot compact corona, warm corona and a cold disk. The schematic geometry is shown in Fig. \ref{fig:KubotaDone}. The model does not yet include all possible complications like the overlap between the phases, and it treats the hot corona as a point-like source in computations of reprocessing of the hard X-rays by the disk but the number of free parameters is large enough to fit the data and to obtain interesting results \cite{NodaDone}. While constructing the model, they suggest that the compact corona luminosity actually saturates at some 2\% of the Eddington luminosity while the source luminosity rises, which automatically leads to a systematic decrease of the role of this component when Eddington luminosity is approached \cite{KubotaDone}. It might be motivated by the fact that compact corona should not reach the compactness parameter leading to efficient pair-creation.

\subsection{UV-luminosity - X-ray luminosity relation}

A very interesting observational tight relation between the UV luminosity and X-ray luminosity has been found by \cite{LussoRisaliti}. Since the relation is non-linear, it can be used in cosmology as a new tool for distance determination and for probing the expansion rate of the Universe \cite{LussoRisaliti2}.

The relation has been found through selection of blue quasars. In this selection, those sources affected by extinction are eliminated but there is still some bias in the mass/accretion rate (SED of quasars with too high mass and relatively low accretion rate would peak in the optical/UV band, e.g. \cite{Czernyetal2011}). Nevertheless, for significant fraction of quasars the relation holds. The same researchers proposed a theoretical explanation which relates the creation of hard X-rays to the transition between the radiation-pressure and gas-pressure domination in the accretion disk \cite{LussoRisaliti2017}. Such a model does not seem convincing to us since in this case the corona must be formally very extended while there are some arguments for compactness of this region. However, we think that the issue is important and perhaps some modification of the model (formation of accreting corona at the transition radius, with dissipation closer in) might offer more convincing explanation. On the other hand, the postulate of the maximum of the hard X-ray luminosity by \cite{KubotaDone} is not identical to the relation proposed by \cite{LussoRisaliti} so further observational work is also needed to set this issue.

\section*{Acknowledgements}

The $\ $ project $\ $ was $\ $ partially $\ $ supported $\ $ by $\ $ National Science Centre, $\ $ Poland, $\ $ grant No. \\ 2017/26/A/ST9/00756 (Maestro 9).

\bigskip
\bigskip
\noindent {\bf DISCUSSION}

\bigskip
\noindent {\bf Elme BREED:} UV/optical variability studies find lags that are much longer than expected from a standard Shakura-Sunyaev discs. Do these new models try to address that?

\bigskip
\noindent {\bf Bozena CZERNY:} Not directly. However, in my opinion most of the effect comes from the fact that only part of the disk radiation comes directly to the observer, while part is being scattered in the disk wind or in the BLR intercloud medium. Such scattering does not change the spectrum but causes additional time delay. We plan to model that in detail with Chris Done.

\end{document}